\documentclass[aps,prl,reprint,superscriptaddress,twocolumn]{revtex4}

\usepackage{amsfonts}
\usepackage{amsmath}
\usepackage[normalem]{ulem}
\usepackage{bm}
\usepackage{graphicx}
\usepackage{subfig}
\usepackage{lipsum}
\usepackage{array,multirow}
\usepackage[utf8x]{inputenc}

\newcommand{\ket}[1]{\vert#1\rangle}

\def\opone{\leavevmode\hbox{\small1\kern-3.8pt\normalsize1}}
\usepackage{color}

\begin{document}
	
	\title{
		Entanglement and non-locality between disparate solid-state quantum memories mediated by photons}
	
	\author{Marcel.li Grimau Puigibert}	
	\altaffiliation{These authors contributed equally to this work.}
	\affiliation{Institute for Quantum Science and Technology, and Department of Physics \& Astronomy, University of Calgary, 2500 University Drive NW, Calgary, Alberta, T2N 1N4, Canada}
	\affiliation{University of Basel, Klingelbergstrasse 82, CH-4056 Basel, Switzerland}
	%-----------------------------
	\author{Mohsen Falamarzi Askarani}
	\altaffiliation{These authors contributed equally to this work.}
	\affiliation{Institute for Quantum Science and Technology, and Department of Physics \& Astronomy, University of Calgary, 2500 University Drive NW, Calgary, Alberta, T2N 1N4, Canada}
	\affiliation{QuTech and Kavli Institute of Nanoscience, Delft University of Technology, 2600 GA Delft, The Netherlands}
	%-----------------------------
	\author{Jacob H. Davidson}
	\affiliation{Institute for Quantum Science and Technology, and Department of Physics \& Astronomy, University of Calgary, 2500 University Drive NW, Calgary, Alberta, T2N 1N4, Canada}
	\affiliation{QuTech and Kavli Institute of Nanoscience, Delft University of Technology, 2600 GA Delft, The Netherlands}
	%-----------------------------
	\author{Varun B. Verma}
	\affiliation{National Institute of Standards and Technology,
		325 Broadway, Boulder, Colorado 80305, USA}
	%-----------------------------
	\author{Matthew D. Shaw}
	\affiliation{Jet Propulsion Laboratory, California Institute of Technology, 4800 Oak Grove Drive, Pasadena, California 91109, USA}
	%-----------------------------
	\author{Sae Woo Nam}
	\affiliation{National Institute of Standards and Technology,
		325 Broadway, Boulder, Colorado 80305, USA}
	%-----------------------------
	\author{Thomas Lutz}
	\affiliation{Institute for Quantum Science and Technology, and Department of Physics \& Astronomy, University of Calgary, 2500 University Drive NW, Calgary, Alberta, T2N 1N4, Canada}
	\affiliation{ETH Z{\"u}rich, Otto-Stern-Weg 1, 8093 Z{\"u}rich, Switzerland}
	%-----------------------------
	\author{Gustavo C. Amaral}
	\altaffiliation{These authors contributed equally to this work.}
	\affiliation{Institute for Quantum Science and Technology, and Department of Physics \& Astronomy, University of Calgary, 2500 University Drive NW, Calgary, Alberta, T2N 1N4, Canada}
	\affiliation{QuTech and Kavli Institute of Nanoscience, Delft University of Technology, 2600 GA Delft, The Netherlands}
	%-----------------------------
	\author{Daniel Oblak}
	\affiliation{Institute for Quantum Science and Technology, and Department of Physics \& Astronomy, University of Calgary, 2500 University Drive NW, Calgary, Alberta, T2N 1N4, Canada}
	%-----------------------------
	\author{Wolfgang Tittel}
	\altaffiliation{Correspondence and requests for materials should be addressed to W. Tittel (email: \mbox{w.tittel@tudelft.nl).}}
	\affiliation{Institute for Quantum Science and Technology, and Department of Physics \& Astronomy, University of Calgary, 2500 University Drive NW, Calgary, Alberta, T2N 1N4, Canada}
	\affiliation{QuTech and Kavli Institute of Nanoscience, Delft University of Technology, 2600 GA Delft, The Netherlands} 
	%\date{\today}
	
	%-----------------------------	

	\begin{abstract}
		Entangling quantum systems with different characteristics through the exchange of photons is a prerequisite for building future quantum networks. Proving the presence of entanglement between quantum memories for light working at different wavelengths furthers this goal. Here, we report on a series of experiments with a thulium-doped crystal, serving as a quantum memory for 794 nm photons,  an erbium-doped fibre, serving as a quantum memory for telecommunication-wavelength photons at 1535 nm, and a source of photon pairs created via spontaneous parametric down-conversion. Characterizing the photons after re-emission from the two memories, we find non-classical correlations with a cross-correlation coefficient of $g^{(2)}_{12} = 53\pm8$; entanglement preserving storage with input-output fidelity of $\mathcal{F}_{IO}\approx93\pm2\%$; and non-locality featuring a violation of the Clauser-Horne-Shimony-Holt Bell-inequality with $S= 2.6\pm0.2$. Our proof-of-principle experiment shows that entanglement persists while propagating through different solid-state quantum memories operating at different wavelengths.
	\end{abstract}
	%203 words in abstract
	
	\maketitle
	
	%\emph{Introduction.}
	Entanglement is central to applications of quantum mechanics \cite{horodecki2009Entanglementreview}. In particular, photon-mediated distribution of entanglement over different and widely spaced quantum systems underpins the creation of a future quantum network \cite{wehner2018quantum}.  Various materials, devices and protocols are currently being studied towards this end \cite{lvovsky2009optical,eisaman2011invited,somaschi2016near}, but, so far, there is no certainty about which elements will constitute its fundamental building blocks. However, it appears likely that they will operate within different wavelength regions, ranging from visible \cite{hensen2015loophole,hedges2010efficient} via near-infrared \cite{2013entanglementRb,yu2019entanglement,saglamyurek2011broadband,clausen2011Entangled,hanschke2018quantum,krutyanskiy2019light}, to telecommunication wavelengths \cite{saglamyurek2015FiberENt,kim2016two}. This will allow leveraging the best properties of each device, and thereby offer heightened capabilities compared to a network consisting of identical quantum systems. 
	
	The development of such networks therefore creates a need, immediately and in the future, for hybridization experiments to bridge existing frequency and bandwidth mismatches. One example is linking quantum memories for light operating at different wavelengths through the exchange of photons. However, so far, only a few investigations have been reported \cite{Rempe2011BEC800nm,2017Nv13C,Riedmatten2017Nature}, and none of them has included a quantum memory functioning at telecommunication wavelength. 
	In this work, we demonstrate entanglement between two atomic frequency comb (AFC)-based quantum memories \cite{afzelius2009AFC} in ensembles of cryogenically-cooled rare-earth ions, one for 794 nm and one for 1535 nm wavelength photons. The first memory employs a thulium-doped lithium-niobate (Tm$^{3+}$:LiNbO$_3$) crystal, the second an erbium-doped fibre (Er$^{3+}$:SiO$_2$). Entanglement is created through the interaction with entangled photons created by spontaneous parametric down-conversion. Both memories allow buffering and re-emitting multiplexed quantum data in feed-forward-controlled spectral or temporal modes, either of which makes them suitable for quantum repeaters \cite{sinclair2014spectral,simon2007rate}. We note that our experiment can also be interpreted as a quantum frequency conversion interface mediated by the spontaneous parametric down-conversion(SPDC) source that connects two quantum memories operating at different wavelengths \cite{Riedmatten2017Nature}.

	\begin{figure*}[t]
		\includegraphics[width=1.00\textwidth]{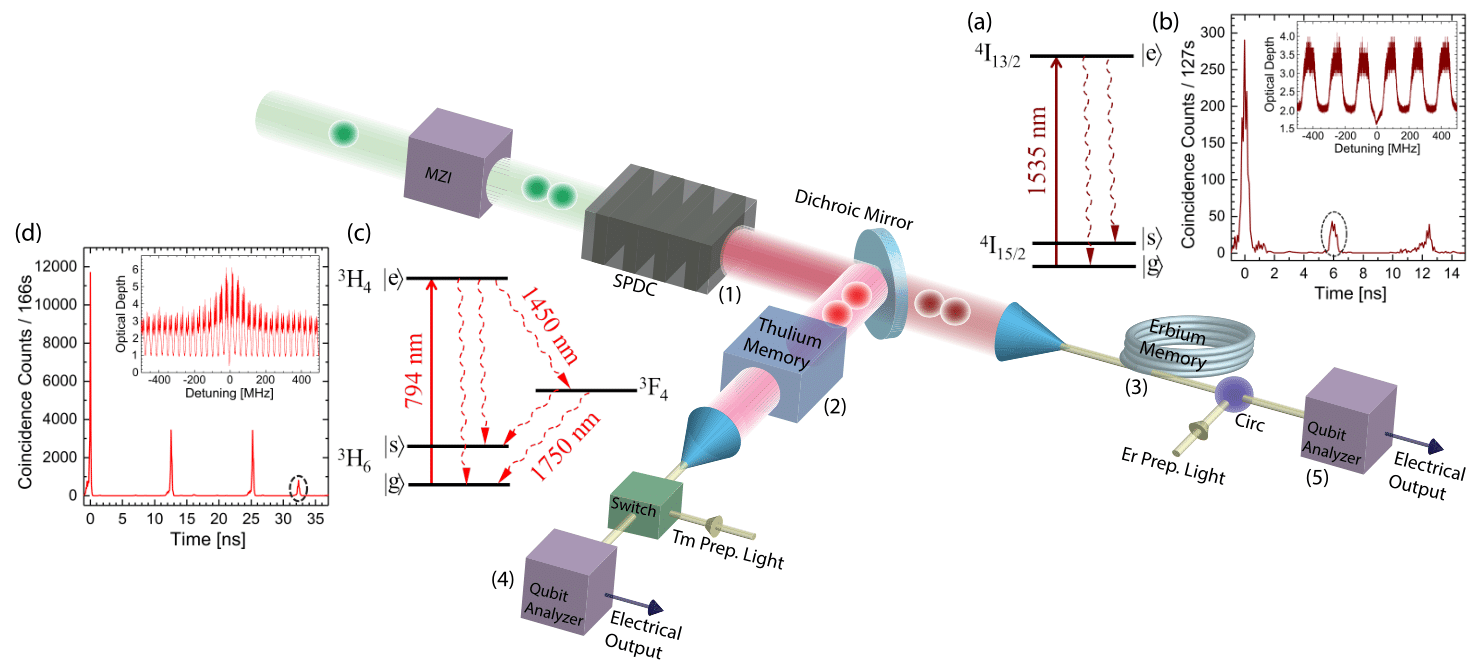}
		
		\caption {Photons from a photon-pair source (1) were directed to two different memories (2, 3), stored, and re-emitted. Analyzers (4, 5) allowed measuring the cross-correlation function, reconstructing density matrices, and testing a Bell inequality. A simplified erbium-level scheme is shown in (a). To create the AFC, erbium ions were frequency-selectively pumped from $\ket{\mathsf{g}}$ via $\ket{\mathsf{e}}$ into the long-lived level $\ket{\mathsf{s}}$. The AFC peak spacing was $\sim$166~MHz (yielding 6~ns storage time) and the total bandwidth was 8~GHz (the inset of (b) shows a 1-GHz-wide section). 
			(b) shows the storage and retrieval of single photons at $\sim$1535~nm wavelength in single temporal modes, heralded by the detection of 794 nm photons post-selected as coincidences with the reference clock signal (see the Supplemental Material). The coincidence peak at 0~ns is due to directly transmitted (non-stored) photons; the one at 6~ns (highlighted by a dashed circle) depicts stored and re-emitted photons. Photons created in subsequent pump laser cycles caused accidental coincidences at multiples of 12.5 ns.  A simplified thulium level scheme is shown in (c). The AFC preparation was similar to that described for erbium. Equivalent measurements to those in (b) are shown in (d), but now for single photons at 794 nm heralded by detections of 1535 nm photons again post-selected. Note the accidental coincidence peaks at 12.5 and 25 ns. The whole AFC frequency range was 10~GHz (a 1-GHz-wide section is depicted in the inset), and the peak spacing $\sim$31~MHz, resulting in 32-ns-long storage.}
		\label{fig:setup}
	\end{figure*}
	
	\emph{Experimental setup.}
	%--------------------------------------------------- unchanged from here
	Our experimental setup, outlined in Fig.~\ref{fig:setup} and further detailed in the Supplemental Material, consisted of four  parts: a source of entangled photon pairs; two solid-state memories for light (one doped with thulium, and one with erbium); and a detection system comprising analyzers (including detectors), coincidence electronics and data processing software. Appropriate configuration of the source and detection system allowed measuring the cross-correlation function, reconstructing density matrices, and testing Bell inequalities with photon pairs before and after storage. 
	
	To create time-bin entangled pairs of photons at 794~nm and 1535~nm wavelengths, short laser pulses at 523~nm and 80~MHz repetition rate (reference clock) were split by an unbalanced Mach-Zehnder (MZ) interferometer into early, $e$, and late, $\ell$, temporal modes, and used to pump a non-linear crystal that is phase-matched for frequency non-degenerate spontaneous parametric down-conversion (SPDC). Assuming the annihilation of exactly one pump photon, the quantum state of the resulting photon pair is described by the so-called $\ket{\phi^+}$ Bell state 
	\begin{equation}
	\ket{\phi^+}=\big (\ket{ee}+\ket{\ell\ell}\big )/\sqrt{2},
	\label{Eq:entangled}
	\end{equation}
	\noindent
	where $\ket{xy} \equiv \ket{x}_\mathrm{794}\otimes\ket{y}_\mathrm{1535}$, with subscripts denoting each photon's wavelength\cite{brendel1999SPDC}. The remaining pump light was removed, and the photons of different wavelengths were separated by a dichroic mirror and sent to separate rare-earth-ion-doped memories. In addition, their spectra were filtered to 8 GHz using an air-spaced etalon (for the 1535 nm photons) and 10 GHz using the inhomogeneously-broadened absorption line of Tm (for the 794 nm photons; see the Supplemental Material for details). We adjusted the paths of the photons such that their storage in the memories was temporally overlapped.  
	
	The Tm$^{3+}$:LiNbO$_3$ crystal, used for storing 794~nm photons, and the Er$^{3+}$:SiO$_2$ fibre, used for storing  1535~nm photons, were cooled to $\sim$0.6~K and exposed to magnetic fields of 125~G  and 1500~G, respectively. Under these conditions it is possible to tailor the inhomogeneously broadened absorption lines of each ensemble of rare-earth ions into a large number of spectrally equidistant  absorption peaks separated by frequency $\Delta$. This feature, known as an atomic frequency comb (AFC), was first described as a quantum memory in \cite{afzelius2009AFC}. 
	After a storage time $\tau=1/\Delta$ the photon is re-emitted from the memory (for a short theoretical description of the storage protocol see the Supplemental Material). In our experiment, we prepared storage times of 32~ns and 6~ns, and memory bandwidths of 10~GHz and 8~GHz for the 794~nm and 1535~nm photons. Note that the short storage times were partly determined by the impossibility to independently optimize the magnetic fields to which the ensembles were exposed in a single cryostat (described in the Supplemental Material).
	
	After re-emission, the photons were directed to analyzers consisting either of a short fibre or an interferometer featuring the same path-length difference as that acting on the pump pulses, and detected using superconducting nanowire single-photon detectors (SNSPDs) cooled to 0.8~K in a second cryostat. Finally, the resulting electronic signals were processed in a time-to-digital converter (TDC), and single-detector count rates as well as time-resolved coincidence count rates were recorded by a computer.

	\emph{Results.}	
	Initially, we characterized the source and the two memories individually. For this, we blocked the long arm of the pump interferometer such that the source produced pairs of photons in a single temporal mode.   
	We measured the system storage efficiencies for heralded single photons for each memory, finding 0.1\% for the Er-doped fibre and 0.4\% for the Tm-doped crystal. Taking independently characterized input and output coupling efficiencies and transmission loss into account, the corresponding device efficiencies were 0.5\% and 2\%. See Fig.~\ref{fig:setup} and Supplemental Material for more information.  
	
	\begin{figure}[t]
		\centering
		\includegraphics[width=0.48\textwidth]{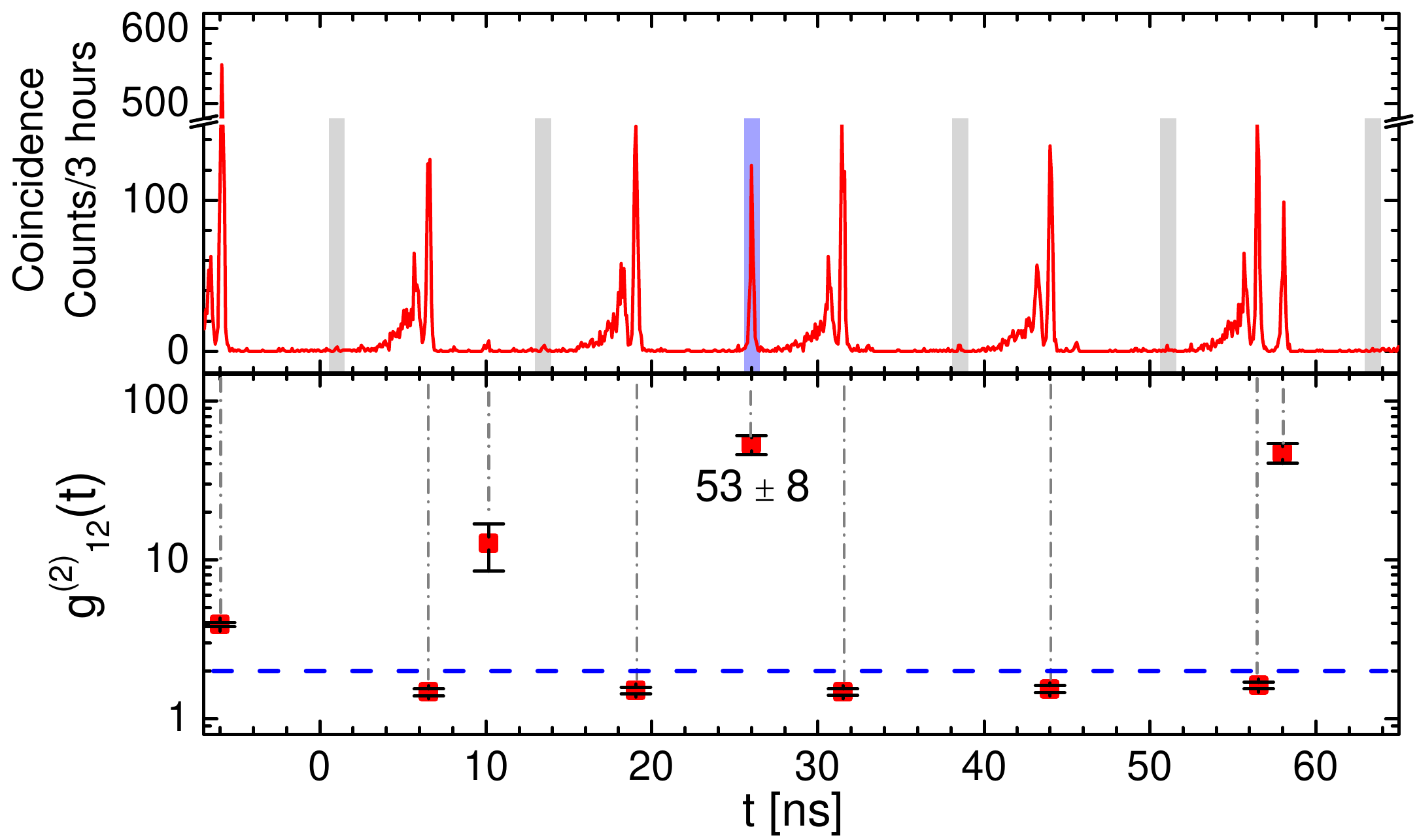}
		\caption{The upper panel shows the coincidence-detection histogram (within a 80 ps bin width) created by starting the TDC with the clock signal ANDed with re-emitted 1535 nm photon detection signals, and stopping it using all 794 nm photons detected after the Tm memory. In the lower panel, the red squares and the black error bars represent, respectively, $g^{(2)}_{12}(\delta t)$ and uncertainties (one standard deviation), calculated from the upper panel histogram. The blue-dashed line shows the maximum classical value of $g^{(2)}_{12}(\delta t)$ = 2. Non-classical correlations between photons after storage can be seen at $\delta t=26$~ns. (The rate R$(26~\textrm{ns})$ is highlighted using a blue bar, and R$(26~\textrm{ns} + \textrm{n}\cdot12.5\textrm{ns})$ using gray bars.) The points at $\delta t=58$~ns and $\delta t=10$~ns were caused by imperfections in the thulium AFC (see the Supplemental Material). \ The data at $\delta t\!=-6~\textrm{ns}$ is due to transmitted (non-stored) 794 nm photons, and data at $\delta t\!=\!-6~\textrm{ns} + \textrm{n}\times 12.5~\textrm{ns}$ corresponds to accidental coincidences with transmitted 794 nm photons emitted in a subsequent pump laser cycle.}
		\label{fig:g2}
	\end{figure}
	
	%---------------------------------------
	
	\begin{figure}[t]
		\centering
		\includegraphics[width=0.30\textwidth]{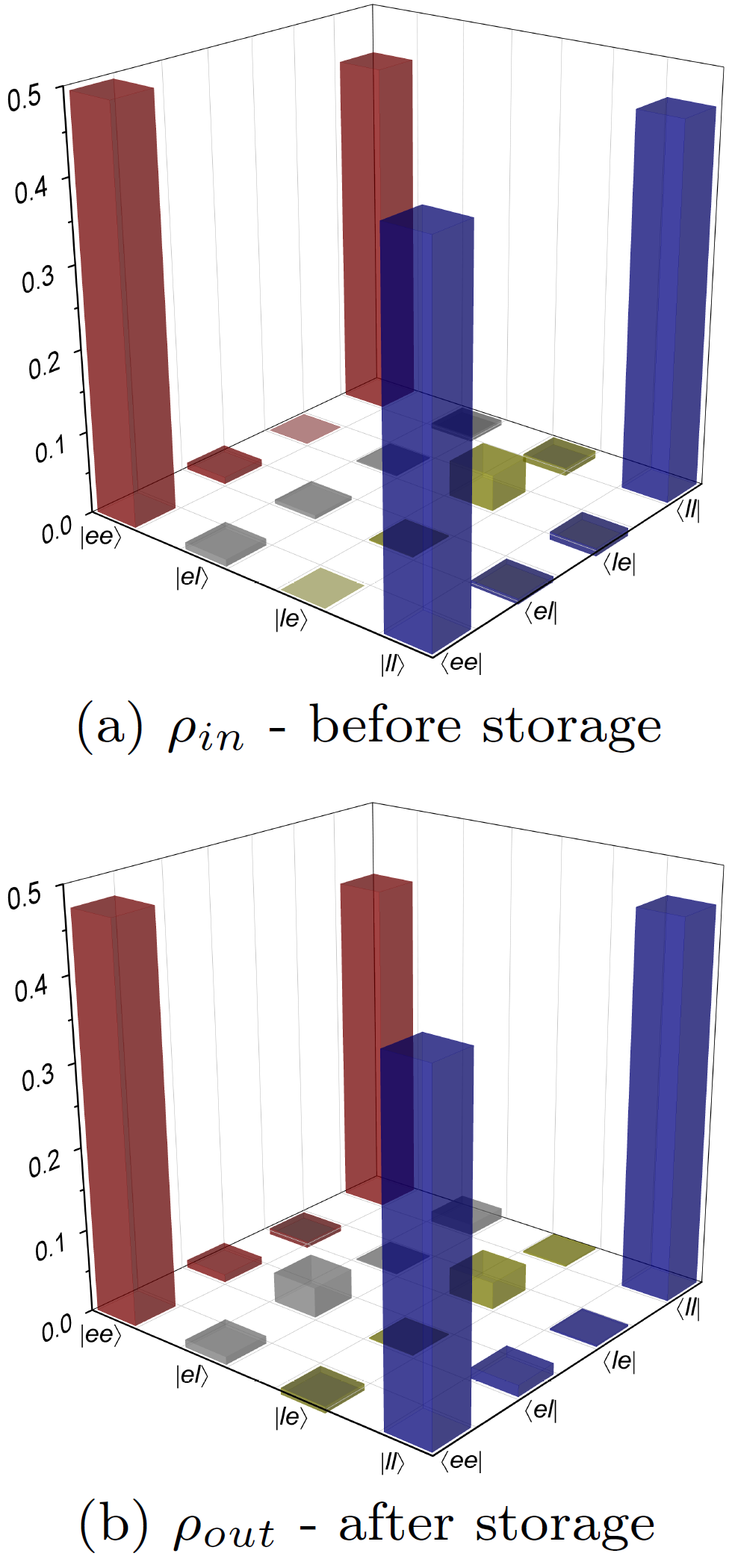}
		\caption{Reconstructed density matrices of entangled time-bin qubits before and after storage. Only the real parts are shown (the absolute values of all imaginary parts are smaller than 0.023).} \label{fig:densMat}
	\end{figure}
	%-----------------------------------------------------------------
	
	\begin{table*}[t]
		\centering
		\vspace{1cm}
		\begin{tabular} {| c |  c | c | c | c |}
			\hline
			& \textbf{Entanglement of formation (\%)} & \textbf{Purity (\%) } & \textbf{Fidelity with $\ket{\phi^{+}}$ (\%)} & \textbf{Input-output fidelity (\%)} \\ 
			\hline
			$\rho_{in}$  & 81.10$\pm$2.23  & 84.57$\pm$1.47 & 91.68$\pm$0.83 & \multirow{2}{*}{93.77$\pm$2.18}\\
			\cline{1-4}
			$\rho_{out}$ & 72.61$\pm$11.70 & 80.14$\pm$7.22 & 87.68$\pm$4.67 &  \\
			\hline
		\end{tabular}
		\caption{
			Values were calculated using the density matrices in Fig.~\ref{fig:densMat}, and uncertainties estimated using Monte-Carlo simulation.}
		\label{tab:entanglement verifications}
	\end{table*}
	
	Then, with the long arm of the pump interferometer still blocked, we measured the cross correlation function $g^{(2)}_{12}(\delta t)$ of the photon pairs before storage in the memories (see the Supplemental Material for more information). 
	Creating photons pairs using the highest pump power available, we found $g^{(2)}_{12}(0)$= 61$\pm$9, which exceeds the maximum value of 2 that is consistent with the assumption of classical fields \cite{ kuzmich2003generation} 
	by 4.5 standard deviations (see Supplemental Material for additional measurements). This verifies the non-classical nature of the photon-pairs. Repeating the measurement with memories in place, we found $g^{(2)}_{12}(\delta t\!=\!32~\textrm{ns}\!-\!6~\textrm{ns}=26~\textrm{ns})$=53$\pm$8 (shown in Fig. \ref{fig:g2}), where 32 ns and 6 ns denote the pre-set storage time of the thulium and erbium memories, respectively. This result demonstrates that non-classical correlations between the members of photon pairs were preserved during storage. 
	
	Next, we opened both arms of the pump interferometer, thereby creating the entangled state described in Eq. \ref{Eq:entangled}. We used two approaches to establish that the photons remain entangled after storage, and hence that we preserved the entanglement during storage in the distinct quantum memories.

	First, we reconstructed the density matrices of the photon pairs before and after storage. This requires measuring a set of joint (bipartite) projectors (see Supplemental Material). Using a maximum likelihood estimation, the density matrices that best fit the measured projection probabilities were reconstructed; the results are shown in Fig \ref{fig:densMat}. In turn, this allowed us to calculate the entanglement of formation, the purity, the fidelity of the measured states with the target state $\ket{\phi^{+}}$ in Eq.~\ref{Eq:entangled}, and the fidelity between input and output states (see Supplemental Material for more information). These figures of merit, summarized in Table \ref{tab:entanglement verifications}, confirm that the photons remained entangled. %%%Therefore, the pair of memories shared an entangled state throughout storage.
	
	Finally, we also performed a test of the CHSH Bell inequality \cite{clauser1969CHSH}, which states that the sum of four correlation coefficients $E(\alpha,\beta)$ is upper-bounded by two if the measurement results can be described by local theories:
	\[
	S=|E(\alpha_1,\beta_1)+E(\alpha_1,\beta_2)+E(\alpha_2,\beta_1)-E(\alpha_2,\beta_2)|\leq 2.
	\label{Eq:CHSH}
	\]
	Here, $\alpha_i$ and $\beta_i$ denote different measurement settings -- in our case phases set by analyzing interferometers. Using the data given in the Supplemental Material, we found S$_{in}$=2.52$\pm$0.02 before, and S$_{out}$=2.6$\pm$0.2 after storage, both of which significantly exceed 2.
	
	Violating the CHSH Bell inequality (or any other Bell inequality) proves non-locality rather than entanglement. However, Bell tests allow certifying entanglement within a device-independent framework, that is without making assumptions about, e.g., the dimensions of the Hilbert spaces describing the individual quantum systems \cite{scarani2012device}. (Note that such certification requires the Bell test to be loophole free, while we made the common assumptions of fair sampling and no signaling.) As such, a Bell-inequality violation is a more stringent test of entanglement than finding positive values for the entanglement of formation.

	\emph{Conclusion and discussion.}
	Our results show that entanglement persists while propagating through different solid-state quantum memories operating at different wavelengths. However, for future use in a quantum network, several factors must be improved. To increase the device efficiency and storage time of the memory for the 794 nm photons, the Tm$^{3+}$:LiNbO$_3$ crystal could be replaced by Tm$^{3+}$:Y$_3$Ga$_5$O$_{12}$ (Tm:YGG)\cite{thiel2014Tm:YGG}, which features better coherence properties, inside an impedance-matched cavity \cite{afzelius2010impedancecavity}. Similarly, the erbium-doped fibre could be replaced by a $^{167}$Er-doped Y$_2$SiO$_5$ crystal (Er:YSO) \cite{Morgan2018coherence}, again inside a cavity. For more information see the Supplemental Material.

	\subsection*{Acknowledgements}  
	The authors thank Erhan Saglamyurek and Neil Sinclair for discussions, and acknowledge funding through Alberta Innovates Technology Futures (AITF), the National Science and Engineering Research Council of Canada (NSERC), and the Netherlands Organization for Scientific Research (NWO). Furthermore, WT acknowledges funding as a Senior Fellow of the Canadian Institute for Advanced Research (CIFAR), and VBV and SWN partial funding for detector development from the Defense Advanced Research Projects Agency (DARPA) Information in a Photon (InPho) program. Part of the detector research was carried out at the Jet Propulsion Laboratory, California Institute of Technology, under a contract with the National Aeronautics and Space Administration (NASA).
	
	%The authors declare that they have no competing financial interests.

	\section*{References}
	
	%\bibliographystyle{naturemag_noURL}
	%\bibliography{Hybrid-ref}

\pagebreak
\widetext
\begin{center}
	\textbf{\large Supplemental Materials: Entanglement and non-locality between disparate solid-state quantum memories mediated by photons}\\

	Marcel.li Grimau Puigibert, Mohsen Falamarzi Askarani, Jacob H. Davidson, Varun B. Verma, Matthew D. Shaw, Sae Woo Nam, Thomas Lutz, Gustavo C. Amaral, Daniel Oblak, Wolfgang Tittel	

\end{center}
%%%%%%%%%% Merge with supplemental materials %%%%%%%%%%
%%%%%%%%%% Prefix a "S" to all equations, figures, tables and reset the counter %%%%%%%%%%
\setcounter{equation}{0}
\setcounter{figure}{0}
\setcounter{table}{0}
\setcounter{page}{1}
\makeatletter
\renewcommand{\theequation}{S\arabic{equation}}
\renewcommand{\thefigure}{S\arabic{figure}}
\renewcommand{\bibnumfmt}[1]{[S#1]}
\renewcommand{\citenumfont}[1]{S#1}

	A detailed schematic of our setup is shown in Fig.~\ref{fig:detailed setup}. In the following we explain the different components.
	
	\subsection{Time-bin entangled photon-pair source}
	
	To generate time-bin entangled photon pairs, we employed a 1047 nm wavelength laser (pump laser) emitting 6 ps-long pulses at an 80 MHz repetition rate. A small part of this signal was sent to a photo-detector that created the reference clock for all measurements. The remaining portion of the pump laser light was directed to a 2~cm-long periodically-poled lithium niobate crystal (PPLN), phase-matched for second-harmonic generation (SHG). The then 18 ps-long pulses, centered at 523.5 nm wavelength, traveled through an unbalanced free-space Mach-Zehnder interferometer (MZI) whose path-length difference corresponded to 1.4~ns travel time difference, thereby creating pulses of light in early and late temporal modes (or time-bins). The emerging pulses were used to pump a second 2~cm long PPLN crystal, where, with small probability of around 1.6\%, spontaneous parametric down conversion (SPDC) occurred, in which a pump photon is annihilated and a photon pair is created in the maximally entangled state $\vert \phi \rangle = \frac{1}{\sqrt{2}} ( \vert e e \rangle + \vert \ell \ell\rangle )$.
	
	As determined by energy conservation and phase-matching, the down-converted photons' wavelengths were centered around 794~nm and 1535~nm -- compatible with our quantum memories. Prior to storage, the 1535~nm photons were filtered to 8 GHz bandwidth using an air-spaced Fabry-Perot cavity (FP) in order to match the bandwidth of the erbium memory. In the case of the 794 nm photons, we took advantage of the fact that the inhomogeneous broadening of Tm in LiNbO$_3$ extends beyond the 10 GHz-wide AFC: photons outside the AFC bandwidth will be absorbed by Tm ions, eventually be spontaneously re-emitted into random directions, and hence have a negligible probability to reach the detector. This enables restricting the photons' bandwidth without the need for an external filter and thereby reduces complexity and loss. Unfortunately, this approach is not possible in the case of erbium because the fibre guides the absorbed and spontaneously emitted photons (outside the AFC's bandwidth) preferentially towards the detector, thus creating noise.
	
	\subsection{Quantum memories}
	
	\subsubsection{Kramers and non-Kramers ions}
	When doped into inorganic crystals, rare-earth elements generally form triply positively (3+) charged ions. Ions with and odd and even number of electrons in their 4f orbital are referred to as Kramers and non-Kramers ions, respectively. Er$^{3+}$ and Tm$^{3+}$ are, in this order, examples for these two classes. In the following, we will discuss the principal differences between Kramers and non-Kramers ions, and their impact on the use as quantum memory and quantum transduction. 
	
	As they have a half-integer spin, Kramers ions feature an unquenched electronic magnetic moment. When doped into crystals, this results in rapid spin-spin relaxation (spin flip-flops) at small magnetic fields, where the spin levels are equally thermally populated. However, by increasing the magnetic field (and hence the electronic Zeeman level splitting), spin-lattice relaxation becomes dominant due to a rapidly growing phonon density of states. Both relaxation mechanisms shorten the electron spin level lifetime, thereby limiting the efficiency of the optical pumping and, as a consequence, the suitability of Kramers ions at small or moderate magnetic fields for AFC-based quantum memory. 
	
	However, at magnetic fields in excess of a few Tesla, only the lowest electronic Zeeman level is thermally populated, resulting in the absence of spin flip-flops. As has been shown recently, it is then possible to use long-lived hyperfine levels (arising from the coupling between nuclear and electronic spins) for efficient optical pumping\cite{Morgan2018coherence}.
	
	Note that the high magnetic-field sensitivity of the electronic Zeeman splitting of Kramers ions  makes them interesting for optical-to-microwave transduction\cite{Brien2014interfacing,fernandez2015transduction} since a transition of a few GHz---the typical level spacing of superconducting qubits---can be achieved by a small external magnetic field. This benefits coherence times of the closely-located superconducting qubits.
	
	\begin{figure*}[t]
		\centering
		\includegraphics[width=0.9\textwidth]{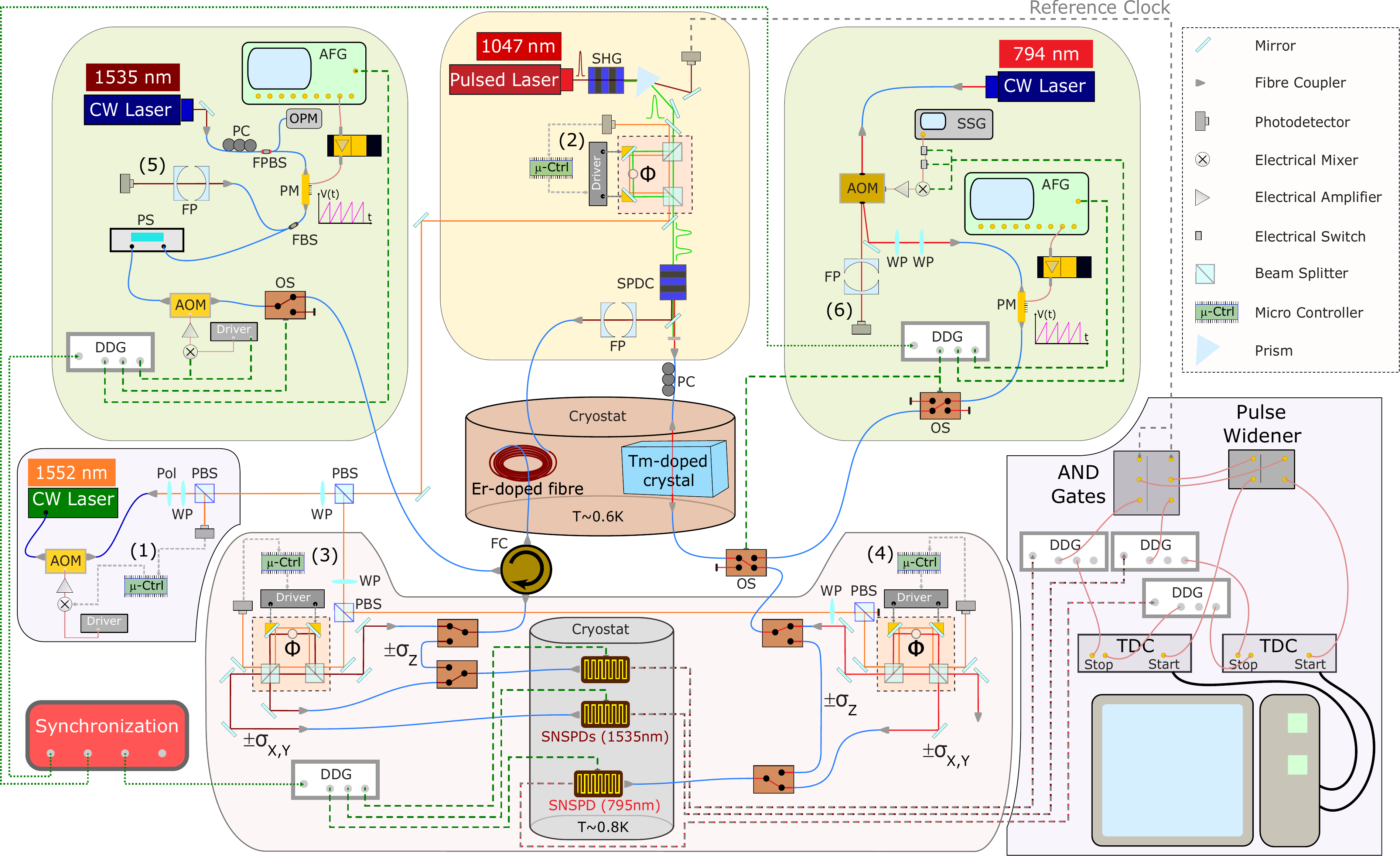}
		\caption{The figure is divided into six panels that compose the four parts described in the main text: top left panel: erbium preparation; top center panel: photon pair source; top right panel: thulium preparation; bottom center panel: measurement, analysis and detection; bottom left panel: stabilization source; bottom right panel: electronics and computer software. Red and dark-red lines denote 794~nm and 1535~nm nm photons traveling in free space, respectively, and orange lines denote stabilization light. The blue wavy lines depict optical fibers for both wavelengths. All other lines depict electronic cables. SSG: synthesized signal generator; Driver: multipurpose signal driver.}
		\label{fig:detailed setup}
	\end{figure*}
	
	Conversely, non-Kramers ions, which do not possess electron spin sub-levels owing to a quenched electronic magnetic moment, appear more suitable for photon-echo-based quantum memory due to their long-lived nuclear Zeeman and hyperfine levels. Their usefulness for optical-to-microwave transduction is limited due to the small magnetic field-dependence of their nuclear Zeeman splittings and hence the need for large magnetic fields to achieve a level splitting of a few GHz.
	
	Note that it is reasonable to assume that the fundamental difference between the two types of rare-earth ions affects the sensitivity of a particular ion’s spectroscopic properties to host imperfections. Hence, a good host crystal for one ion is not necessarily a good one for an ion from the other class.

	\subsubsection{The AFC protocol}
	When a single photon in a well-defined temporal mode is collectively absorbed by ions in an AFC at time $t=0$, the result is a so-called Dicke state of the form
	\begin{equation}
	\ket{\psi}_A=\frac{1}{\sqrt{N}}\sum_{j=1}^N c_j e^{-i2\pi n_j\Delta\cdot t}e^{ik\cdot z}\ket{\mathsf{g}_1,..., \mathsf{e}_j,..., \mathsf{g}_N}.
	\label{Eq:Dicke}
	\end{equation}
	\noindent
	Here, $\mathsf{g}$ denotes the atomic ground state and $\mathsf{e}$ the excited state, $k$ is the wave number and $z$ the propagation direction of the light, and $j=1..N$ labels the ions interacting with the photon with weighted amplitudes $c_j$ ($c_j$ depends on the ion's detuning $n_j \Delta$ with respect to the centre frequency of the photon and on its position within the absorbing medium)\cite{afzelius2009AFC}.

	\subsubsection{Erbium doped fibre - AFC preparation}
	
	Continuous-wave telecommunication-wavelength light at 1535 nm (Er-memory preparation light) was injected from a diode laser into a single-mode optical fibre. A polarization controller (PC) and a fibre-coupled polarization beam splitter (FPBS) with one of its outputs connected to an optical power meter (OPM) allowed controlling the intensity in the second output of the FPBS. The light was then directed to a phase modulator (PM), employed to serrodyne shift its frequency in discrete intervals (each defining a region in which spectral hole-burning took place). The PM was driven by an arbitrary function generator (AFG) connected to an electrical amplifier, allowing for shifts up to $\pm$~10~GHz.
	
	After the PM, the light went through a polarization scrambler (PS) with scrambling frequency of 6 kHz, used to address all erbium ions despite their randomly oriented transition dipole moments (this is caused by the amorphous SiO$_2$ host). In-between the PS and the erbium-doped fibre, an acousto-optic modulator (AOM) was placed. It was driven by 500 ms-long electrical pulses that were synchronized to the master clock. The amplitude of the pulses could be controlled for optimal AFC preparation.  
	In addition, we also inserted a latch-mode optical switch (OS) and a fibre-pigtailed circulator (FC). The OS was connected to port \#1 of the FC while port \#2 was connected to the Er-doped fibre, and port \#3 to the analyzers.
	
	The Er-doped fibre was exposed to a 1500 G magnetic field, which lifts the degeneracy of the electronic ground state and creates two long-lived Zeeman levels with second-long lifetimes.  
	
	The interaction between the modulated pump light and the erbium ions leads to frequency-selective persistent spectral hole burning, and, after 1000 repetitions of the 500 $\mu$s-long burning sequence, to an 8-GHz-wide AFC (see Fig.~\ref{fig:detailed setup} in the main text for a 1 GHz-wide central section). After a wait time of 200 ms, included to ensure the decay of ions from the excited level (avoiding spontaneously emitted photons during the subsequent step), we repeatedly sent, stored (for 6 ns), and retrieved members of photon pairs for a total of 700 ms. During this phase of the experimental cycle, the optical switch was toggled to prevent memory preparation light from leaking into the single-photon detectors. See  Fig.~\ref{fig:timeSequence} for a timing diagram.\\
	
	To read the AFC, port \#3 of the FC was connected to a photo-detector followed by an oscilloscope. After the preparation of the AFC, a reading waveform was applied to the PM, which scanned the laser frequency across the prepared spectral structure from -1 to +1 GHz while transmission was monitored. To calibrate the measured optical depth (OD), and hence estimate the AFC efficiency, we recorded the transmitted intensity of another 2 ms-long pulse that burned a narrow spectral section to complete transparency. 
	
	\subsubsection{Thulium-doped lithium niobate crystal - AFC preparation}
	
\begin{figure*}[t]
		\centering
		\includegraphics[width=0.9\linewidth]{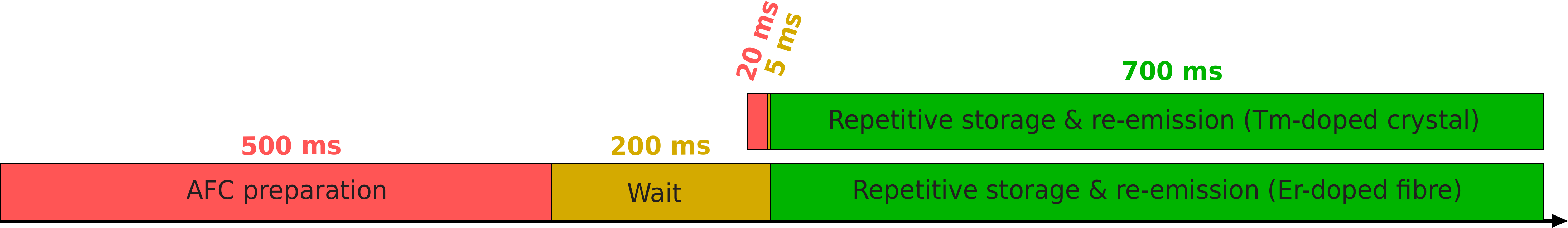}
		\caption{Shown are synchronized preparation, wait and storage periods for the two memories.} 
		\label{fig:timeSequence}
	\end{figure*}

	Visible-wavelength light at 794 nm (Tm-memory preparation light) was generated by a diode laser. To obtain a clean spatial mode, the beam was coupled into and out of a single-mode optical fibre. The light was then directed to a single-pass AOM, creating 20 ms-long pulses, and the deflected first-order beam steered to a set of wave-plates for polarization control before being coupled into a fibre-pigtailed PM. Similar to the Er memory preparation, the PM was driven by an AFG and an electrical amplifier for serrodyne shifting. Each sweep took 1 ms and was repeated 20 times per preparation sequence (see  Fig.~\ref{fig:timeSequence}).
	
	After serrodyne shifting in the PM, the light was directed to two fibre-pigtailed optical switches that allowed for routing collimated memory preparation light and single-photons in and out of the crystal while avoiding leakage of the classical light into the waiting and storage periods. 
	
\begin{figure*}[t]
		\centering
		\includegraphics[width=0.6\linewidth]{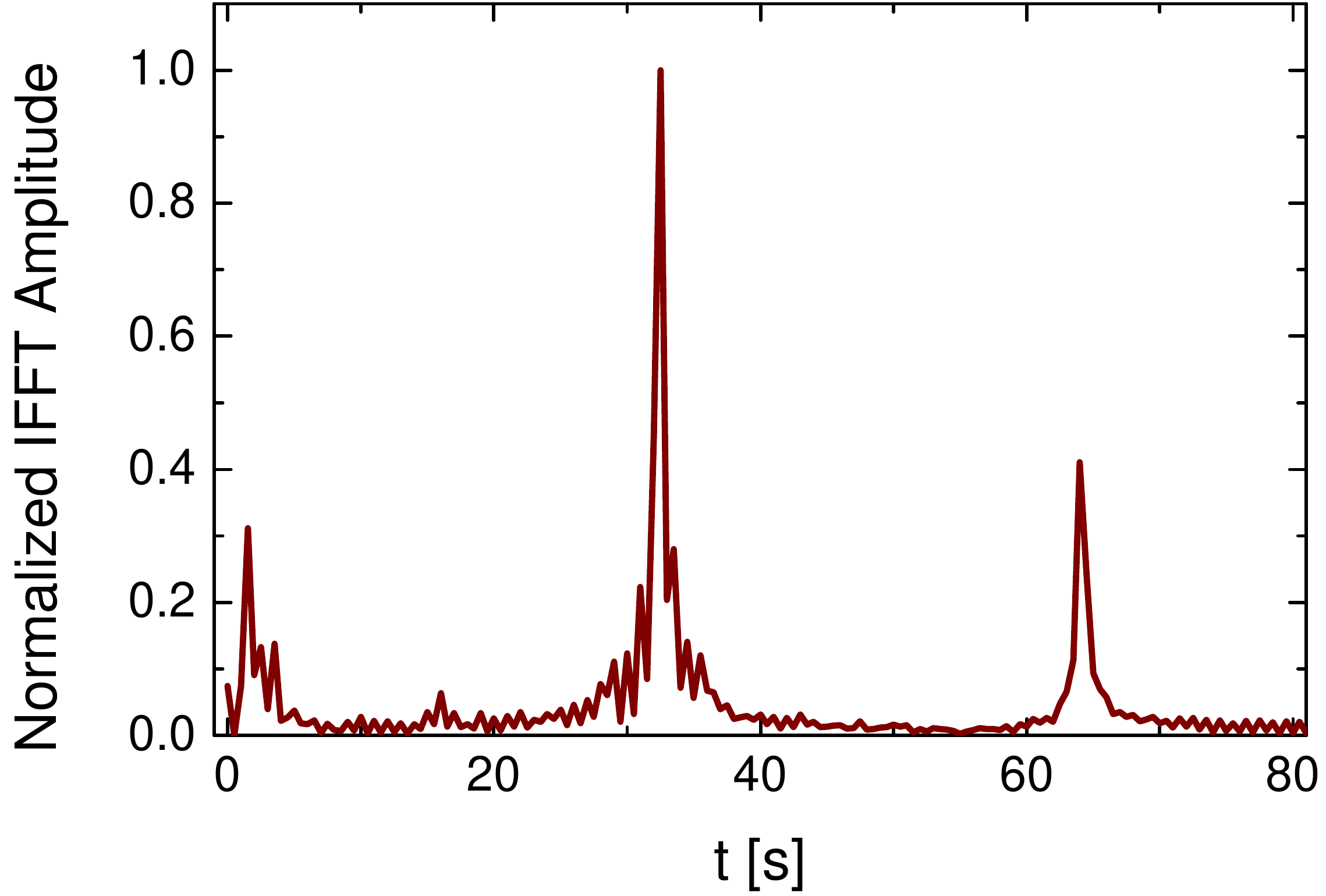}
		\caption{Fourier transform of the Tm AFC spectrum (a 1 GHz-broad section is shown in the inset of Fig.~1d in the main text), showing possible recall times of 16ns, 32 ns and 64 ns.}
		\label{fig:FFT}
	\end{figure*}

	The crystal was cooled to 0.6~K and exposed to a 125~G magnetic field aligned along its $C_3$ axis, resulting in two nuclear Zeeman sub-levels. Due to their long lifetimes, on the order of minutes, the interaction between the preparation light and the thulium ions leads to persistent spectral holes and allowed creating a 10-GHz-wide AFC (see Fig.~1 in the main text for a 1 GHz-large central section). After a waiting time of 5~ms to avoid spontaneous emission noise, a 700~ms-long storage period was started during which many photons were stored and re-emitted 32 ns later. However, due to spurious modulation of the AFC with periods different from $\Delta$=1/32~ns $\sim$31~MHz, there were two additional moments at which partial rephasing occurred, namely after half and twice the intended storage time 1/$\Delta$ (see Fig.~2 in the main text). This is confirmed by the Fourier transform of the measured AFC structure, shown in  Fig.~\ref{fig:FFT}.

	\subsubsection{Optimization of magnetic field} 
	To optimize an AFC for quantum state storage, it is necessary to carefully set the magnetic field to which an ensemble of rare-earth-ions is exposed. In the case of two different ensembles, this implies the need to tune both fields individually. However, in this investigation both fields were created by the same superconducting solenoid (the field difference is due to the ensembles being in different locations with respect to the solenoid), and individual adjustment, and hence optimization of the individual storage efficiencies, was impossible. Instead, we chose a setting that resulted in the best overall efficiency. 
	
	More precisely, the erbium-doped fibre requires a magnetic field of around 1500~G to allow for quantum state storage (at other fields spin relaxation prevents efficient optical pumping, as discussed above).
	
	At this setting, the field at the location of the Tm-crystal is only around 125~G, leading to small nuclear Zeeman splitting. As is always the case if the ground state splitting is smaller than the AFC bandwidth, optical pumping must be tailored to transfer atomic populations from troughs of the AFC to neighboring teeth. Hence, storage time---given by the inverse tooth spacing--- and level splitting are not independent anymore. In addition, such cross-pumping results in a finesse (the ratio between the teeth spacing and the teeth width) of 2, thereby limiting the storage efficiency\cite{afzelius2009AFC}. In short, the storage time and the maximum efficiency of the Tm memory are both determined by the need to expose the Er memory to a particular magnetic field. However, since the ground-state electronic Zeeman splitting of erbium exceeds the AFC bandwidth, the storage time in the erbium memory is only limited by the coherence time -- it is not determined by the level splitting.
	
	\subsubsection{System and device efficiencies}
	
	The analysis of the device efficiencies of the two memories was conducted based on the following two methods. First, by comparing the detection rate of the recalled photons at the output of the cryostat with that of all photons input into the cryostat, we calculated the system efficiency: $\eta_{\textrm{system}}=R_{out}/R_{in}$. In addition to the actual probability for absorbing and re-emitting photons---the device efficiency--- the system efficiency also includes fibre transmission and, for the Tm memory, coupling in and out of the crystal. To estimate the latter two (jointly referred-to as $\eta_{\textrm{coupling}}$), we burned a fully transparent spectral hole in the absorption profile of the rare-earth ions using strong light, and then evaluated the ratio of output power over input power: $\eta_{\textrm{coupling}}=P_{out}/P_{in}$. The device efficiency can then be calculated using $\eta_{\textrm{device}}=\eta_{\textrm{system}}/\eta_{\textrm{coupling}}$.
	
	In the case of the erbium-doped fiber, the input count rate was 417 Hz, and the rate of retrieved photons behind the cryostat was 1.8 Hz (both rates were measured in coincidence with the heralding photon). This translates approximately into a 0.1\% system efficiency. Taking into account a 20\% coupling efficiency, we find a device efficiency of 0.5\%. 
	For the thulium-doped crystal quantum memory, identical measurements resulted in 0.4\% for the system and 2.0\% for the device efficiency.
	
	Alternatively, we estimated the device efficiency through the analysis of the AFC spectral profile, which we obtained as described in the \textit{AFC preparation} section of the Methods (see Fig. 1 in the main text for two examples). The values of the optical depth of the remaining background due to imperfect optical pumping ($d_0$) and the optical depth of the teeth ($d_1$), as well as the Finesse (F) was inferred by fitting a train of Gaussian functions to the acquired profiles. Then, by using\cite{afzelius2009AFC}  
	\begin{equation}
	\eta_{device} = (d_{1}/F)^2 e^{-d_{1}/F} e^{-7/F_{2}} e^{-d_{0}},
	\end{equation}
	we calculated the device efficiency. We found 0.5\% for the erbium memory and 2\% for the thulium memory, consistent with the values estimated using the first method.
	
	\subsubsection{Optimization of memory wavelengths} 
	
	The coincidence count rates in our experiments depend on several interrelated factors, which have to be optimized jointly: the need for energy correlations between the photons belonging to each pair, and the wavelength-dependent storage efficiencies in both memories. To optimize the rate, we measured the device efficiencies as a function of the wavelengths of the two photons produced by the source. The results are presented in  Table~\ref{tab:wavelenthVSeff}.
	
	\subsection{Control and stabilization}
	
	The experimental setup depicted in  Fig.~\ref{fig:detailed setup} contains many individual elements that must be controlled to ensure physical parameter stability---in particular wavelengths, intensities and phases---throughout the measurements. Four feedback loops were used, allowing for intensity stabilization of the laser used for phase stabilization of various interferometers ((1) in Fig.~\ref{fig:detailed setup}); phase stabilization of the pump interferometer ((2) in Fig.~\ref{fig:detailed setup}); and phase stabilization of the two analyzing interferometers ((3) and (4) in  Fig.~\ref{fig:detailed setup}). Furthermore, to assure that the wavelengths of the lasers used for memory preparation were stable, wavelength control was required ((5) and (6) in Fig.~\ref{fig:detailed setup}).

	\begin{figure*}[t]
		\centering
		\includegraphics[width=0.6\linewidth]{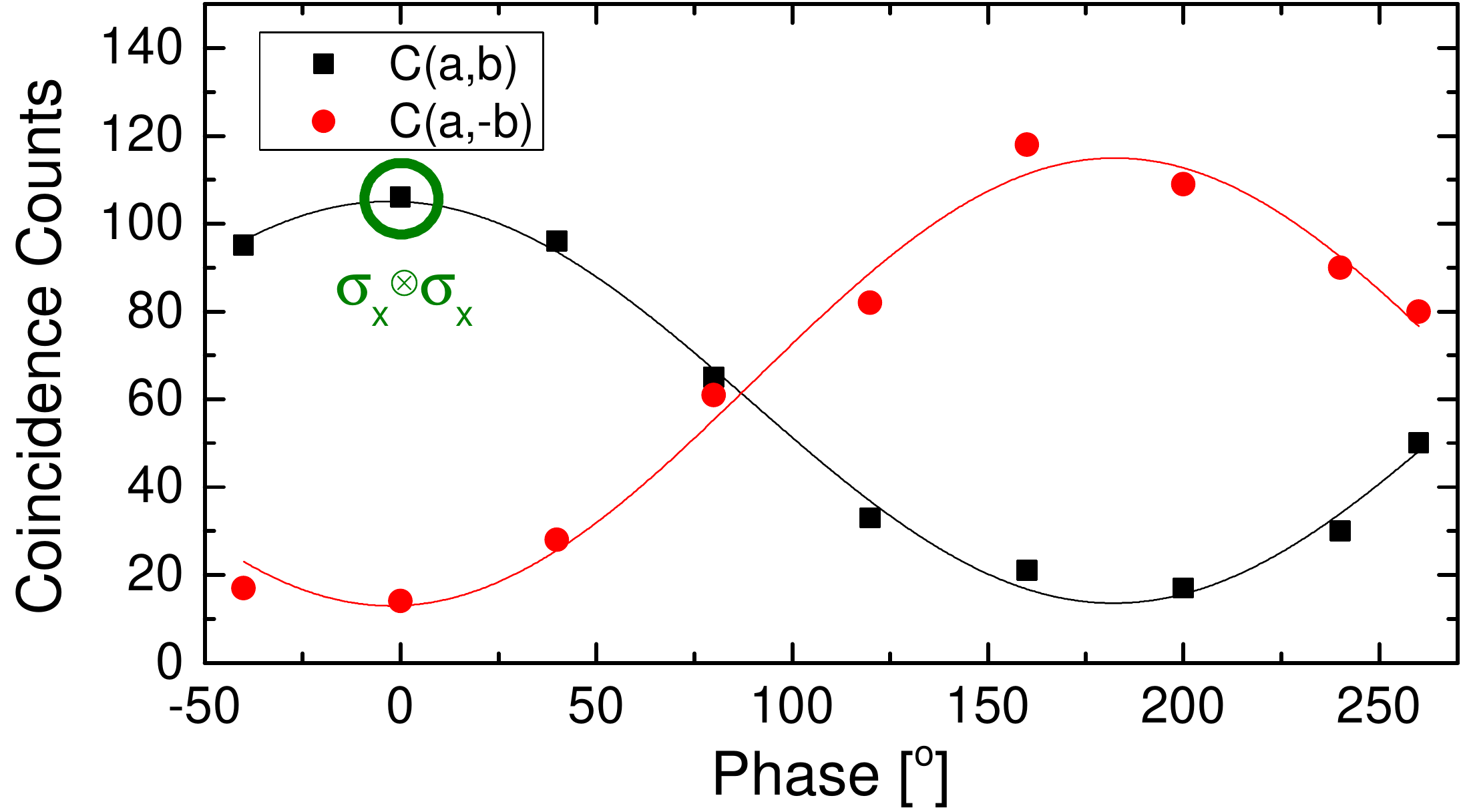}
		\caption{Visibility curve taken by fixing the phases of the pump and 1535 nm interferometers to 0, and sweeping the 794 nm interferometer's piezo voltage, i.e. phase. The y-axis shows coincidences between the clock signal, and detections of the re-emitted 794 nm and transmitted 1535 nm photons. The accumulation time per data point was 60 seconds. The set of phases that define the $\sigma_x\otimes\sigma_x$ projection corresponds to the point of the curve where C(a,b) is maximum and  C(a,-b)  minimum.}
		\label{fig:visCurve}
	\end{figure*}
	
	\subsubsection{Intensity Stabilization -- Feedback Loop 1}
	
	Continuous-wave light was generated by a fibre-coupled and internally frequency-locked 1532 nm wavelength laser. After passing through a fibre-pigtailed AOM, used to modulate the light intensity, it was collimated and sent through a bulk-optics polarizer, a wave-plate and a PBS.  One output was subsequently used for interferometer stabilization (see below), and the other coupled to a free-space variable-gain photo-detector. To stabilize the intensity of the light in the former output, the detector signal was used as a parameter for the AOM's modulation signal. 
	
	\subsubsection{Phase Stabilization -- Feedback Loops 2-4}
	
	The frequency- and intensity-stabilized light, split into three individually intensity-controlled beams, was injected into and detected behind each interferometer. Due to interference, any phase variation then caused a variation of the light intensity measured by the corresponding detector. In turn, this allowed locking the interferometers to specific phases using piezo-mounted mirrors.
	
	Fig.~\ref{fig:visCurve} displays typical single detection rates (accumulated during 60 seconds per data point) in the case where the pump interferometer and the  interferometer analyzing the 1535 nm photons were fixed, and the phase of the 794~nm interferometer was swept from 0$^o$ to 360$^o$; the set of phases that corresponds to the $\sigma_x\otimes\sigma_x$ projector is highlighted.
	
	\subsubsection{Laser stability monitoring -- circuits 5 and 6}
	
	To ensure wavelength stability of the 1535 nm memory preparation laser, we collected a fraction of the emitted light using a fiber beam splitter (FBS) inserted behind the PM. This light was then directed to a Fabry-Perot (FP) optical cavity, detected by a photo-detector, and monitored on an oscilloscope. This allowed us to observe fluctuations of the laser's wavelength. The wavelength of the 794 nm preparation laser was similarly monitored by another FP cavity.  
	
	\subsection{Data collection and analysis}
	
	\subsubsection{Photon detection and data collection}
	
	The 794~nm and 1535~nm photons were detected using tungsten silicide (WSi) superconducting nanowire single-photon detectors (SNSPDs) cooled inside a sorption fridge to $\sim$0.8K. Each detector's efficiency was optimized for the wavelength of the detected photons\cite{redaelli2016design}. The average photon-detection efficiency, including fibre loss inside the cryostat, was measured to be 70~$\%$ for all detectors. The temporal jitter of all detectors was around 250 ps, and the dark count rates below 100~Hz.  To adjust arrival times of detection signals at the TDC, digital delay generators (DDGs) were placed at the outputs of the SNSPDs (see Fig.~\ref{fig:detailed setup}), and to avoid  detections of leaked preparation light, the detectors were only enabled during the storage period.
	
	As the 80 MHz clock rate was too high to start the TDC, we instead used a down-sampled version generated from an AND gate that additionally received signals from 1535 nm photon detections. All non-relevant start signals, i.e. starts that were not associated with photon pair detections, were discarded.
	
	\subsubsection{Acquisition of correlations} 
	
\begin{figure*}[t]
		\centering
		\includegraphics[width=0.6\linewidth]{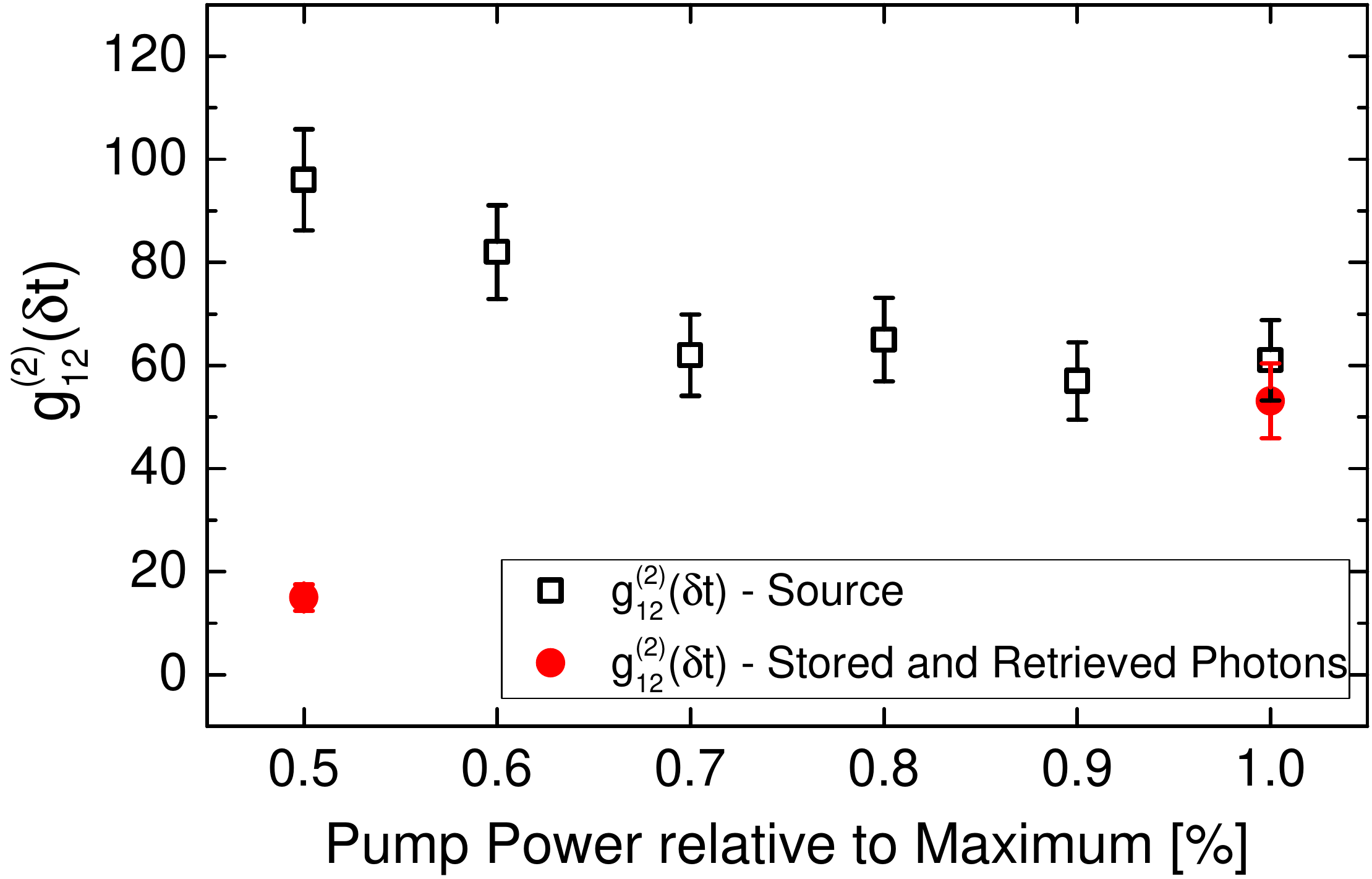}
		\caption{Measured $g^{(2)}_{12}\left(\delta t\right)$ for different values of the pump power. As expected, the value of $g^{(2)}_{12}\left(\delta t\right)$  for the source increases with decreasing pump power. However, to achieve a higher SNR, we used the highest available pump power throughout all measurements involving quantum memories.}
		\label{fig:g2all}
	\end{figure*}

	The second-order cross correlation function is defined as $
	g^{(2)}_{12}(\delta t)=P_{12}(\delta t)/P_1P_2$,
	where $P_1$ and $P_2$ denote the probabilities of individual detections, and $P_{12}(\delta t)$ is the coincidence detection probability for events occurring with time difference $\delta t$.  
	Assuming ``true" coincidences between photons from the same pair and before storage to arise at $\delta t\!=\!0$, ``accidental" coincidences between photons from subsequent pairs happen at $\delta t' = \textrm{n}\cdot 12.5$~ns ($\textrm{n}\in\mathbb{Z}\neq0$), with 12.5~ns being the pump laser's inverse repetition rate. Since the photon-pair generation process is spontaneous, there are no statistical correlations between subsequently emitted pairs. This allows us to express $g_{12}^{(2)}(0)$ using coincidence count rates R$(0)$ $\propto P_{12}(0)$ and $\langle$R$(\textrm{n}\cdot 12.5\textrm{ns})\rangle_\textrm{n}\propto P_1P_2$\cite{kuzmich2003generation}, where ``$\langle..\rangle_\textrm{n}$" denotes averaging over n
	: $g_{12}^{(2)}(0)$=R$(0)$/$\langle$R$(\textrm{n}\cdot 12.5\textrm{ns})\rangle_\textrm{n}$.
	More generally, 
	\begin{equation}
	g_{12}^{(2)}(\delta t)=\textrm{R}(\delta t)/\langle\textrm{R}(\delta t+12.5\textrm{ns})\rangle_\textrm{n} .
	\label{Eq:cross}
	\end{equation}
	In Fig.~\ref{fig:g2all}, the values of $g_{12}^{(2)}(\delta t)$ for the source and for the stored and retrieved photons is presented as a function of the pump power.

	The histogram presented in Fig. 1 of the main text, which depicts time-dependent detections of photons stored and retrieved from the erbium (thulium) memory, was acquired by starting the TDC with the clock down-sampled by detections of heralding 794nm (1535nm) photons generated by the SPDC source. For all other measurements ($g^{(2)}_{12}(\delta t)$, quantum state tomography, and Bell inequality violation), the TDC was always started by the clock after down-sampling by detections of stored and re-emitted 1535 nm photons. To calculate the cross correlation coefficient from experimental data using Eq.\ref{Eq:cross}, we varied n between -5 and +5.

	\subsubsection{Qubit analysis}
	
	Free-space interferometers with path-length differences identical to that of the pump interferometer and controlled relative phases were used to project individual qubits of the entangled state onto superposition states $\frac{1}{\sqrt{2}} ( \vert e \rangle + e^{i\theta} \vert \ell \rangle $, where the phase $\theta$ was established as described in Fig.~\ref{fig:visCurve}. Apart from the active phase stabilization described above, the phase of the interferometers was passively stabilized by temperature controlling its enclosure. Alternatively, to measure qubits in the canonical bases, i.e. $ \vert e \rangle$ and $ \vert \ell \rangle $, the interferometers were replaced by a short fibre, and the photon arrival time was recorded.
	
	To reconstruct two-photon density matrices, we measured several bipartite projectors composed of individual projections onto eigenstates of combinations of the Pauli operators $\sigma_x$, $\sigma_y$ and $\sigma_z$. In the case of temporal mode encoding, this is done by means of interferometers with suitably-chosen phases (for $\sigma_x$ and $\sigma_y$), or delay lines (for $\sigma_z$), respectively.

	\subsubsection{Entanglement of formation, purity and fidelity calculation}  
	The entanglement of formation is defined as 
	\begin{equation}
	E_{F}(\rho) = H \Big( 0.5+0.5\sqrt{1-C^{2}(\rho)}\Big),
	\label{Eq:EoF}
	\end{equation}
	where $H(x)= - x \ log_{2}(x) - (1-x) \ log_{2}(1-x)$ and C($\rho$) is the concurrence, which is defined as 
	\[
	C(\rho) = \max\{ 0, \lambda_{1}-\lambda_{2}-\lambda_{3}-\lambda_{4} \}.
	\]
	The $\lambda_{i}$'s are the eigenvalues of the reconstructed density matrix shown in Fig. 3 of the main text. 
	
	The fidelity between $\rho$ and $\sigma$ is 
	\begin{equation}
	F(\rho , \sigma) = \Bigg( \textrm{Tr} \Big( \sqrt{\sqrt{\rho}\sigma \sqrt{\rho}} \Big) \Bigg)^{2}
	\label{Eq:fidelity}
	\end{equation}
	and the purity of a state $\rho$ is 
	\begin{equation}
	P = \textrm{Tr}(\rho^{2}).
	\label{Eq:purity}
	\end{equation}

	\subsubsection{Bell-inequality test}  
	
	The four correlation coefficients that compose S in Eq. 2  of the main text were measured as
	\begin{equation}
	E\left(a,b\right)~=~\frac{C\left(a,b\right)-C\left(a,-b\right)}{C\left(a,b\right)+C\left(a,-b\right)}
	\end{equation}
	using projectors $a = \sigma_x$, $a' = \sigma_y$, $b = \left(\sigma_x + \sigma_y\right)$, and $b' = \left(\sigma_x - \sigma_y\right)$. As explained in more detail in \cite{saglamyurek2011broadband}, these projections correspond to detecting photons in specific outputs of interferometers with appropriately chosen phases. Tables \ref{tab:density matrix} and \ref{tab:Bell inequality} present the individual values acquired for Quantum State Tomography and CHSH-Bell inequality tests, respectively.
	
	%----------------------------------------------------------
	
	\begin{table*}[t]
		\centering
		\vspace{0.2cm}
		\begin{tabular} {| c | c | c | c | c |}
			\hline
			$\lambda_\mathrm{signal}$ [nm] & $\lambda_\mathrm{idler}$ [nm] & $\eta_{\textsc{Tm}}$ & $\eta_{\textsc{Er}}$ & $\eta_{\textsc{Tm}}\times\eta_{\textsc{Er}}$ \\ 
			\hline
			795.15 & 1534.05 & $0.93\%$ & $0.50\%$ & $0.0095\%$ \\
			\hline                              
			794.85 & 1535.17 & $1.87\%$ & $0.48\%$ & $0.0182\%$ \\
			\hline                              
			794.68 & 1535.80 & $2.00\%$ & $0.50\%$ & $0.0200\%$ \\
			\hline  
			794.55 & 1565.29 & $1.49\%$ & $0.49\%$ & $0.0148\%$\\
			\hline
		\end{tabular}
		\caption{Efficiency versus wavelength for pairs of wavelengths available in the SPDC source.}
		\label{tab:wavelenthVSeff}
	\end{table*}
	
	\newpage
	
	\begin{table*}[t]
		\centering
		%\vspace{0.5cm}
		\begin{tabular} {|c || c | c | c | c | c | c|}
			\hline
			& $\sigma_z\otimes\sigma_z$  & $\sigma_z\otimes\sigma_x$ & $\sigma_z\otimes\sigma_y$ & $\sigma_x\otimes\sigma_z$ & $\sigma_x\otimes(\sigma_x+\sigma_y)$ & $\sigma_x\otimes(\sigma_x-\sigma_y)$\\
			\hline
			P$_{\textrm{in}}$ \% & 49.55$\pm$2.48 & 25.51$\pm$0.43 & 24.96$\pm$0.43 & 26.85$\pm$0.41 & 41.72$\pm$1.08 & 40.15$\pm$1.06\\
			P$_{\textrm{out}}$ \% & 48.21$\pm$2.48 & 25.00$\pm$5.89 & 24.49$\pm$7.07 & 25.00$\pm$5.10 & 40.91$\pm$6.82 & 42.50$\pm$10.31\\
			\hline
			\hline
			& $\sigma_x\otimes\sigma_y$ & $\sigma_y\otimes\sigma_z$& $\sigma_y\otimes\sigma_x$ & $\sigma_y\otimes(\sigma_x+\sigma_y)$ & $\sigma_y\otimes(\sigma_x-\sigma_y)$ & \\
			\hline
			P$_{\textrm{in}}$ \% & 23.82$\pm$0.57 & 27.15$\pm$0.42 & 25.22$\pm$0.59 & 8.66$\pm$0.58 & 40.39$\pm$1.05 & \\
			P$_{\textrm{out}}$ \% & 22.22$\pm$5.56 & 25.26$\pm$5.16 & 26.39$\pm$6.05 & 6.82$\pm$3.94 & 40.00$\pm$8.16 & \\
			\hline
			\hline
			& $\sigma_z\otimes\sigma_{-z}$  & $\sigma_z\otimes\sigma_{-x}$ & $\sigma_z\otimes\sigma_{-y}$ & $\sigma_x\otimes\sigma_{-z}$ & $\sigma_x\otimes(\sigma_{-x}+\sigma_{-y})$ & $\sigma_x\otimes(\sigma_{-x}-\sigma_{-y})$\\
			\hline
			P$_{\textrm{in}}$ \% & 2.26$\pm$0.10 & 23.37$\pm$0.42 & 23.63$\pm$0.41 & 20.81$\pm$0.36 & 9.04$\pm$0.50 & 9.85$\pm$0.53\\
			P$_{\textrm{out}}$ \% & 2.30$\pm$0.54 & 26.39$\pm$6.05 & 26.53$\pm$7.36 & 27.08$\pm$5.31 & 9.09$\pm$3.21 & 7.50$\pm$4.33\\
			\hline
			\hline
			& $\sigma_x\otimes\sigma_{-y}$ & $\sigma_y\otimes\sigma_{-z}$& $\sigma_y\otimes\sigma_{-x}$ & $\sigma_y\otimes(\sigma_{-x}+\sigma_{-y})$ & $\sigma_y\otimes(\sigma_{-x}-\sigma_{-y})$ & \\
			\hline
			P$_{\textrm{in}}$ \% & 26.18$\pm$0.60 & 20.48$\pm$0.36 & 24.78$\pm$0.59 & 41.34$\pm$1.26 & 9.61$\pm$0.51 & \\
			P$_{\textrm{out}}$ \% & 27.78$\pm$6.21 & 24.21$\pm$5.05 & 23.61$\pm$5.73 & 43.18$\pm$9.91 & 10.00$\pm$4.08 & \\
			\hline
			\hline
			& $\sigma_{-z}\otimes\sigma_{z}$ & $\sigma_{-z}\otimes\sigma_{x}$& $\sigma_{-z}\otimes\sigma_{y}$ & $\sigma_{-x}\otimes\sigma_{z}$ & $\sigma_{-y}\otimes\sigma_{z}$ & \\
			\hline
			P$_{\textrm{in}}$ \% & 2.15$\pm$0.11 & 24.45$\pm$0.43 & 24.74$\pm$0.42 & 28.16$\pm$0.42 & 28.54$\pm$0.43 & \\
			P$_{\textrm{out}}$ \% & 2.17$\pm$0.53 & 23.61$\pm$5.73 & 24.49$\pm$7.07 & 26.04$\pm$5.21 & 25.26$\pm$5.16 & \\
			\hline
			\hline
			& $\sigma_{-z}\otimes\sigma_{-z}$ & $\sigma_{-z}\otimes\sigma_{-x}$& $\sigma_{-z}\otimes\sigma_{-y}$ & $\sigma_{-x}\otimes\sigma_{-z}$ & $\sigma_{-y}\otimes\sigma_{-z}$ & \\
			\hline
			P$_{\textrm{in}}$ \% & 46.03$\pm$0.41 & 26.66$\pm$0.44 & 26.67$\pm$0.44 & 24.18$\pm$0.39 & 23.83$\pm$0.39 & \\
			P$_{\textrm{out}}$ \% & 47.31$\pm$2.46 & 25.00$\pm$5.89 & 24.49$\pm$7.07 & 21.88$\pm$4.77 & 25.26$\pm$5.16 & \\
			\hline
		\end{tabular}
		\caption{Joint-detection probabilities used for the reconstruction of the density matrices before (in) and after (out) storage.}
		\label{tab:density matrix}
	\end{table*}
	
	\begin{table*}[t]
		\centering
		\vspace{0.2cm}
		\begin{tabular} {| c | c | c | c | c | }
			\hline
			& a $\otimes$ b & a $\otimes$ $b'$ & $a'$ $\otimes$ b & $a'$ $\otimes$ $b'$ \\ 
			\hline
			E$_{in}$ \% & 60.59$\pm$1.34  & 64.39$\pm$1.27 & -61.56$\pm$1.31 & 65.40$\pm$1.48  \\
			\hline                              
			E$_{out}$ \% & 68.73$\pm$11.29  & 63.03$\pm$8.22 & -59.28$\pm$10.32 & 68.07$\pm$10.34 \\
			\hline
		\end{tabular}
		\caption{Correlation coefficients used to test the CHSH-Bell inequality.}
		\label{tab:Bell inequality}
	\end{table*}

	\subsection{Future improvements}
	
	In order to meet the benchmarks imposed by a spectrally-multiplexed quantum repeater\cite{sinclair2014spectral}, the storage time and storage efficiency have to be significantly improved for both the Tm- as well as the Er-doped memory.  
	
	The simplest implementation that allows one to achieve in principle unit storage efficiency is the use of an impedance-matched cavity\cite{afzelius2010impedancecavity} in which the reflectivity of the front mirror is determined by the round-trip loss in the rare-earth-ion doped crystal (the back mirror is assumed to be 100\% reflective). This idea has very rapidly enabled increasing the quantum memory efficiency from a few percent to 56\%  \cite{sabooni2013efficient}. In addition, the cavity also restricts the AFC bandwidth to the cavity linewidth, and spectral modes to resonances spaced by the cavity free-spectral range. Calculations based on reasonable parameters for crystal length (e.g. 1 mm) and optical depth (e.g. two) result in an AFC width on the order of 1 GHz and memory channels spaced by around 5 GHz. These are values that allow the use of our current photon pair source without modification.

	The optical coherence time---the fundamental limit to the storage in optical coherence---of the 795 nm transition in Tm:Y$_3$Ga$_5$O$_{12}$ (Tm:YGG) at 1.2~K approaches 500 $\mu$s, a factor of 15 more than in Tm:LiNbO$_3$ at the same temperature, with further improvements expected at lower temperature\cite{thiel2014Tm:YGG}. Paired with long-lived nuclear Zeeman levels, allowing for efficient optical pumping, and a simple sub-level structure, this makes this crystal a very promising memory candidate for a quantum repeater architecture based on spectral multiplexing\cite{sinclair2014spectral}. In particular, when combined with an impedance-matched cavity, quantum state storage with 90\% efficiency during a time on the order of 100 $\mu$sec -- sufficient for an elementary link length in excess of 20 km -- can be expected.
	
	Er-doped Y$_2$SiO$_5$ (Er:YSO) features the longest optical coherence time of an impurity in a solid-state material, around 4.4~ms. This is an improvement by more than a factor of 10$^{4}$ compared to Er:SiO$_2$. However, as mentioned in the \textit{Kramers and non-Kramers ions} section of the Methods, due to significant spin-spin and spin-lattice relaxation, a quantum memory has so far not been demonstrated in this crystal. But it has very recently been discovered that the application of a 6~T magnetic field across a $^{167}$Er:Y$_2$SiO$_5$ crystal results in a hyperfine sub-level lifetime exceeding 1 min and an optical pumping efficiency of 95\%, making it very promising for storage of telecommunication wavelength photons\cite{Morgan2018coherence}. Hence, quantum state storage with 90\% efficiency during a ms using optically-excited coherence should be possible. Moreover, instead of increasing the hyperfine sub-level lifetime in order to make AFC-based quantum memories possible, one could also reduce the excited level lifetime by means of the Purcell effect in a nano-cavity\cite{zhong2017purcell}.
	
	Note that by replacing the currently used host materials LiNbO$_3$ and SiO$_2$ by YGG and YSO, no significant changes to the preparation procedure of the quantum memories are necessary. Furthermore, adding an impedance-matched cavity to the memory adds the requirement for proper mode-matching, which, however, constitutes no significant technical problem. It should also be noted that the spectral multiplexing scheme proposed in \cite{sinclair2014spectral} does not rely on on-demand recall -- reemission after a predetermined storage time (given by the inverse AFC teeth spacing), is sufficient. In short, our proof-of-principle demonstration can readily be translated into workable quantum technology by employing optimized host materials and suitable cavities.

\end{document}